\documentstyle[times,balanced,epsf,prl,aps]{revtex}

\newcommand{\BE}{\begin{equation}}
\newcommand{\EE}{\end{equation}}
\newcommand{\BC}{\begin{center}}
\newcommand{\EC}{\end{center}}
\newcommand{\BI}{\begin{itemize}}
\newcommand{\EI}{\end{itemize}}
\newcommand{\BA}{\begin{eqnarray}}
\newcommand{\EA}{\end{eqnarray}}

\begin{document} 

\draft
\tightenlines

\title{
Average Patterns of Spatiotemporal Chaos: A Boundary Effect}

\author{
V. M. Egu\'\i luz$^{a,}$\cite{email}, 
P. Alstr{\o}m$^{b}$, E. Hern\'andez-Garc\'\i a$^{a}$ and O. Piro$^{a}$} 
\address{
$^{a}$Instituto Mediterr\'aneo de Estudios Avanzados 
IMEDEA\cite{imedea} (CSIC-UIB),
E-07071 Palma de Mallorca (Spain) \\
$^{b}$Center for Chaos and Turbulence Studies (CATS),
The Niels Bohr Institute, Blegdamsvej 17, 2100 Copenhagen {\O}, Denmark\\
} 

\date{\large{{\sl Published in} Phys. Rev. {\bf E 59}, 2822-2825 (1999)}}

\maketitle

\begin{abstract}  

Chaotic pattern dynamics in many experimental systems show structured
time averages. We suggest that simple universal boundary effects underly this
phenomenon and exemplify them with the Kuramoto-Sivashinsky equation
in a finite domain. As in the experiments, averaged patterns in the 
equation recover global symmetries locally broken in the chaotic field. 
Plateus in the average pattern wavenumber as a function of
the system size are observed and studied and the different 
behaviors at the central and boundary regions are discussed.
Finally, the structure strenght of
average patterns is investigated as a function of system size.

\end{abstract} 


\begin{twocolumns}

Experimental studies of the chaotic pattern dynamics in Faraday waves
\cite{Gluckman93,Bosch}, in rotating thermal convection
\cite{Ning}, and in electroconvection \cite{Rudroff}, reveal that
spatio-temporal complex patterns can have surprisingly ordered
time averages. The form of these average patterns (square, circular,
hexagonal) is determined by the underlying symmetry
\cite{symm,pere} imposed by the boundary conditions. 
Although the instantaneous patterns fluctuate chaotically,
they are biased towards the average pattern because they have
short-lived patches spatially in phase with this average.
This phase rigidity seems to come from the boundaries, and
quantization effects appear due to the finite size of the container.
The amplitude of the time-averaged pattern 
depends on the system size and control parameters.
It is strongest near the sidewalls, and decays with increasing
distance from the sidewalls and with increasing fluctuations
about the ordered averaged state. For very large containers the ordered
average pattern exists only near the sidewalls.

Given the general features of average patterns suggested
by experiments, it seems surprising that their possible existence  
and characterization have not been addressed within 
the standard model equations displaying spatiotemporal
chaotic states \cite{CrossHohenberg}. One possible reason for this is that 
periodic boundary conditions are usually considered in theoretical studies.  
In such situation spatial translational invariance homogenizes out 
any time average (unless some unexpected ergodicity 
breaking takes place). Boundary conditions breaking translational symmetry, 
as in the experiments, are thus needed to obtain nontrivial average patterns. 
Motivated by this fact,
we here consider the Kuramoto-Sivashinsky equation, one of the 
prototype equations showing spatiotemporal
chaos, in bounded one and two
dimensional domains. We show that ordered average patterns do appear, despite 
the strong fluctuations, and we discuss the universal aspects of
wave number selection and amplitude variations. More directly, our
analysis of average patterns may be relevant and suggestive for
experiments on phase turbulence in convection cells \cite{Kuramoto},
fluids flowing down an inclined wall \cite{Sivashinsky80},
and flame front propagation \cite{Sivashinsky83,Gorman}.

The Kuramoto-Sivashinsky equation \cite{Kuramoto,Sivashinsky77} is perhaps
the simplest partial differential equation exhibiting spatio-temporal chaos. 
The equation in one dimension has the form: 
\begin{equation}     
h_{t}= - h_{xx} - h_{xxxx} + ( h_{x})^2~,
\label{hks}   
\end{equation}    
where $h=h(x,t)$ is a real function, $x\in [0,L]$, and the
subscripts stand for derivatives. In two dimensions the spatial
derivative is replaced by a gradient and the second derivative
by a Laplacian. The only control parameter
for the equation is the length of the domain $L$; prefactors to the
terms in Eq.~(\ref{hks}) can be scaled out.  An equivalent equation 
for $u=h_x$ can be obtained by taking the derivative of Eq.~(\ref{hks})
with respect to $x$, 
\begin{equation}     
u_{t}= - u_{xx} - u_{xxxx} + 2 u u_{x}~.
\label{uks}   
\end{equation}    
Eq.~(\ref{hks}) possesses 
translational symmetries ($h \to h + h_0$, $x \to x + x_0$)  
a reflexion symmetry ($h \to h$, $x \to - x$), and an infinitesimal Galilean
symmetry ($x \to x + 2 v t$, $h \to h + v x$). Eq.~(\ref{uks}) is also
invariant under translations ($u \to u $, $x \to x + x_0$) and under 
a Galilean symmetry ($x \to x + 2 v t$, $u \to u + v$). A different 
reflexion symmetry
is valid in this case ($u \to - u$, $x \to - x$). 

The stability of the laminar solution $h=0$ ($u=0$) is analyzed by
linearizing Eq.~(\ref{hks}) [Eq.~(\ref{uks})]. For commonly used boundary
conditions the 
growth rate $\mu$ for the Fourier mode of wavenumber $k$
is $\mu=k^2 -k^4$. In two dimensions $k^2$ is replaced by $|{\bf k}|^2$.
The laminar solution is unstable for all modes
within $0<k<1$. The fastest growing mode has a wavenumber $k_0=1/\sqrt{2}$ 
corresponding to a wavelength $\lambda_0 =2 \sqrt{2} \pi \approx 8.9$.
The wavelength $\lambda_0$ serves as a basic length scale, and the
system size $L$ is naturally measured in units of this scale,
$L/\lambda_0$, which is called the aspect ratio.
Beyond the linear range, the nonlinear term becomes important and produces 
growth (linear in time) of the mean 
value of $h$, while the mean value of $u$ saturates.
For $L$ large enough to permit a sufficient number of unstable Fourier
modes, the solution exhibits spatio-temporal chaotic behavior that
can be associated with a disordered evolution of a cellular pattern.

Many studies have been devoted to the bulk behavior of the
Kuramoto-Sivashinsky system \cite{CrossHohenberg}. In relation
to average patterns however the boundaries are of paramount importance, as
discussed above. 
Here, we consider two types of boundary conditions. One of them
is the rigid boundary conditions, where 
\BE
\label{Ru}
u(0,t) = u(L,t) = u_x(0,t) = u_x(L,t) = 0~,
\EE
or equivalently,
\BE
\label{Rh}
h_x(0,t) = h_x(L,t) = h_{xx}(0,t) = h_{xx}(L,t) = 0~. 
\EE
Our other choice of boundary conditions is
\BE
\label{SFu}
u(0,t) = u(L,t) = u_{xx}(0,t) = u_{xx}(L,t) = 0~,
\EE
or equivalently, 
\BE
\label{SFh}
h_x(0,t) = h_x(L,t) = h_{xxx}(0,t) = h_{xxx}(L,t) = 0~,
\EE
which we call stress-free boundary conditions, with reference to
similar conditions in hydrodynamics.

We integrate the Kuramoto-Sivashinsky equation using explicit 
finite-differences of first order 
in time, second order in space for the linear terms, and fourth order in space
for the nonlinear term. 
The time step is chosen sufficiently small
to avoid any spurious behavior. 
The number of grid points used is $128$ in one dimensional simulations
and  $64 \times 64$ in two dimensions. 
In all cases, the simulations were started from random initial conditions. 

\begin{figure} 
\begin{center}
\def\epsfsize#1#2{0.3\textwidth}
\leavevmode
\epsffile{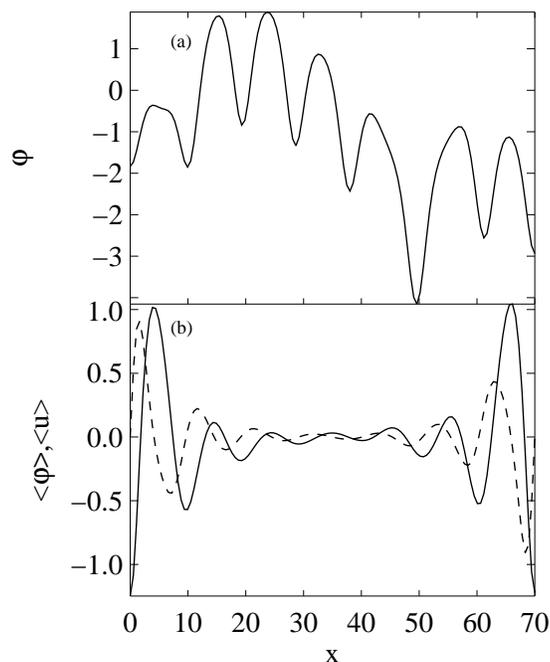} 
\end{center}
\vspace{.6cm}
\caption{\label{StressFree} Results from simulations of the one-dimensional
Kuramoto-Sivashinsky equation with stress-free boundary conditions. System 
size $L=70$. 
a) A characteristic front $\varphi$ at time $t=1000$. 
b) The time average of the front $\varphi$ (continuous line) and of $u$ 
(dashed).}
\end{figure}

 We are interested in the average pattern of $u$ and the average pattern
of the front $\varphi$ of $h$,
\BE
\varphi(x,t) = h(x,t) - \frac{1}{L}\int_0^L h(x,t) dx~.
\label{defm}
\EE
To optimize the measurements of the average, the sampling was first 
started well beyond the initial transient behavior.  
For the system sizes considered, the typical transient
time was limited to approximately $20$ time units, 
and we discarded the first $100$ time units. Then, averages where taken from
configurations sampled every $5$ time units. A total of $10,000$ configurations 
per run were included, and further average over 10 runs with independent
random initial conditions was performed. This is a large sample, but was
necessary to compensate for the slow convergence of the averages produced by 
the long-range time correlations present in the KS equation. 

\begin{figure} 
\vspace{-.4cm}
\begin{center}
\def\epsfsize#1#2{0.3\textwidth}
\leavevmode
\epsffile{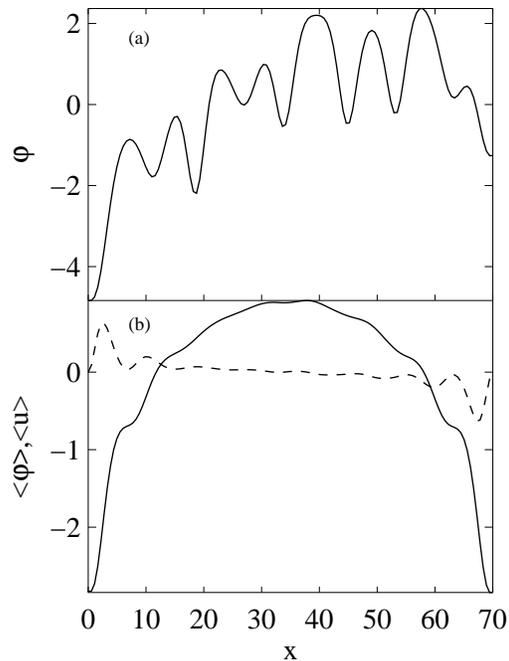} 
\end{center}
\vspace{.4cm}
\caption{\label{Rigid} The same as Fig.~\protect{\ref{StressFree}} but for 
rigid boundary conditions}.
\end{figure}

\vspace{-.6cm}
In both one and two dimensions and with both rigid and stress-free
boundary conditions we obtain non-trivial and ordered time-averaged
patterns from the spatio-temporally chaotic evolution 
(Figs.~\ref{StressFree}-\ref{twosizes}), emphasizing
that the formation of average patterns in spatio-temporal complex
systems is general despite the presence of very large
fluctuations. The presence of boundaries breaks the translational symmetry of
the equations. The boundary conditions (\ref{Ru})-(\ref{SFh}) respect however 
its reflection symmetries (for reflexions with respect to the center of the
domain). As in the experiments, here we find that the 
average patterns display these remaining symmetries. In the two-dimensional
case the average pattern recovers also the square symmetry of the integration
domain (Fig.~\ref{2d}).  
Except for the one-dimensional case with
stress-free boundary conditions (Fig.~\ref{StressFree}), an overall parabolic 
profile of the average front is obtained (see Figs.~\ref{Rigid}-\ref{2dcut}). 
For the derivative $u$ a mean slope is obtained \cite{Zaleski}. 
This parabolic profile is a peculiarity of the
Kuramoto-Sivashinsky equation. It can be removed by considering 
the second derivative of the front $\varphi$ instead of $\varphi$ itself; for 
this 
variable (and for the 
Laplacian in two dimensions) the discussion for all the cases is very similar 
to the 
one-dimensional stress-free situation, that we address in further detail 
in the remaining of the paper. 

 Fig.~\ref{twosizes} shows the average patterns for $L=60$ and $L=100$. The 
number of oscillations increases with 
the size of the system, although only those close to the boundaries are large. 
Furthermore, the distance between consecutive 
maxima is close (but not equal, see below) to the characteristic length scale
$\lambda_0\approx 8.9$. 
Similar observations were done in the experiments referred to at the
beginning. In Fig.~\ref{wavenumbers} the number of local maxima $N$ in the
average front    
is shown for increasing values of the aspect ratio $L/\lambda_0$. $N$ is 
written in terms of the average distance $\lambda$ 
between two consecutive minima, $N=L/\lambda$. The line where
$\lambda=\lambda_0$ is also indicated in Fig.~\ref{wavenumbers}.  
Plateaus are obtained at every integer $N$ between $6$ and $11$ 
for system sizes $L$ between $59$ and $112$. 
The average distance $\lambda$ is consistently larger than $\lambda_0$. 
Consistent deviations (positive or negative) are also known from
the Faraday wave experiments \cite{Gluckman93}. 

\begin{figure} 
\begin{center}
\def\epsfsize#1#2{0.4\textwidth}
\leavevmode
\epsffile{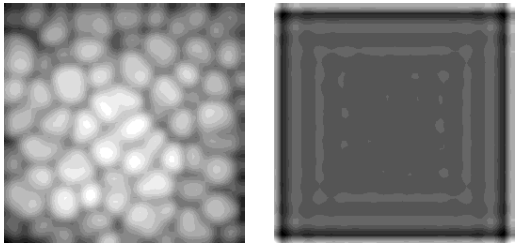}
\end{center}
\caption{\label{2d} Results from simulations of the two-dimensional
Kuramoto-Sivashinsky equation with stress-free boundary conditions. 
System size $L\times L=70\times 70$. Left: A characteristic instantaneous 
image of the front $\varphi$ ($t=1000$), 
different values of $\varphi$ are coded with different gray levels. Right: The
average of $\nabla^2 \varphi$, showing the square symmetry. The average of 
the Laplacian instead of the front itself is shown to eliminate the dominant
parabolic shape of the average, thus improving the visibility of the sidewall 
oscillations. 
}
\end{figure}

If only the central region is considered, 
the plateaus fall off. More specifically, consider the
`central region', defined as the domain ranging from the second local
minimum to the second last local minimum of $\varphi$ (see Fig.~\ref{twosizes}). 
The rest of the the pattern is thus considered the `boundary region'. 
We now determine the average distance 
$\lambda_c$ between consecutive minima in the central region for 
various system sizes $L$, and find the number of maxima $L/\lambda_c$ 
characteristic 
for the central region. The results are shown in Fig.~\ref{wavenumbers}. 
Intriguingly, the plateaus now fall off, an observation also done
in experimental studies of the central region \cite{Gluckman93}. 
In order to explain this `fall off' effect, we determine
the average distance $\lambda_b$ between minima in the boundary regions. 
Over the entire range of system sizes considered, these 
distances changes very little, not more than 4\%, so that to a first
approximation we can consider $\lambda_b$ independent of $L$. 
For the central region we now have
\begin{equation}
\frac{L}{\lambda_c} = (N-4)\frac{L}{L-4\lambda_b} \simeq
N + 4 \left( \frac{(N-4)\lambda_b}{L} - 1 \right)~.
\end{equation}
The last approximation is valid for $\lambda_b/L$ small. 
For $\lambda_b$ constant, it is seen that $L/\lambda_c$ falls off
as $\sim L^{-1}$ within a given plateau characterized by $N$.
Thus an almost constant value of
$\lambda_b$ serves as a generic explanation for the generally observed
fall off of the plateaus. The overall picture is that when $L$ is increased 
the total number of oscillations tends to remain constant, as well as
$\lambda_b$, so that $\lambda_c$ increases. This situation continues until 
the local wavelength in the central region is far enough from $\lambda_0$,
moment at which a new oscillation is accommodated and a jump in $N$ occurs.  

\begin{figure} 
\begin{center}
\def\epsfsize#1#2{0.29\textwidth}
\leavevmode
\epsffile{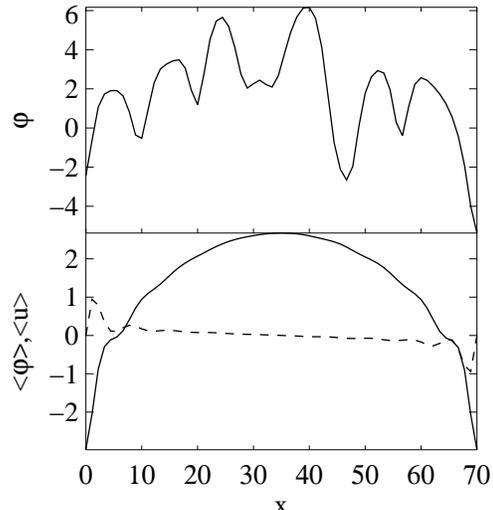}
\end{center}
\vspace{.3cm}
\caption{\label{2dcut} 
a) A central horizontal cut of the configuration shown in the left panel of
Fig.~\protect{\ref{2d}}. b) Solid: a central horizontal cut of the time 
average of the front  $\varphi$ for the same parameter values; dashed: 
the same cut for the time average of $\varphi_x$. }
\end{figure}
 
From Fig.~\ref{twosizes} it is clear that the amplitude $A(x)$ of the
average pattern in general decays with increasing distance from the
boundaries \cite{Egolf}. Experimental studies show 
the same behavior \cite{Gluckman93,Bosch,Ning,Rudroff}. 
To quantify this observation, we consider the
spatial average $\left<A^2\right>=L^{-1}\int_0^L A^2 dx$. The variation of
$A_{rms}=\sqrt{\left< A^2 \right>}$ with system size is shown in 
Fig.~\ref{Arms}, showing a power-law dependence as $A_{rms}\sim L^{-1/2}$. 
We explain this fact by noting that Fig.~\ref{twosizes} indicates  
that $A_{rms}$ receives its largest contribution from the
boundaries, so that the integral in the definition of 
$\left<A^2\right>$ becomes a
constant for system sizes larger than the boundary region. Thus the factor 
$L^{-1}$ in the definition of $\left<A^2\right>$ becomes the dominant 
$L$-dependence thus
providing the observed behavior of $A_{rms}$. 

In conclusion, we have established the formation of ordered time-averaged
patterns in the Kuramoto-Sivashinsky equation, in one and two dimensions,
and with rigid as well as stress-free boundary conditions. The average pattern
recovers the symmetries which are respected by both the equation and the
boundary conditions. The amplitude
is strongest at the boundaries and decays with increasing distance to them. 
The law of decay has been found and explained. 
We have determined the selected wavelength $\lambda$, its
variation with system size $L$, and interpreted the different behavior between 
the central and boundary regions. Most of these observations are also found in
experimental systems for which the Kuramoto-Sivashinsky equation does not
apply, thus indicating its generic, mainly geometrical, origin: What is 
relevant for these phenomena to occur is the 
occurrence of strong enough chaotic fluctuations in the presence 
of non-trivial boundaries. 

We acknowledge the financial support of the Spanish Direcci\'on General de
Investigaci\'on Cient\'\i fica y T\'ecnica, contract numbers PB94-1167 and
PB94-1172.


\end{twocolumns}
\begin{twocolumns}

\begin{figure} 
\begin{center}
\def\epsfsize#1#2{0.46\textwidth}
\leavevmode
\epsffile{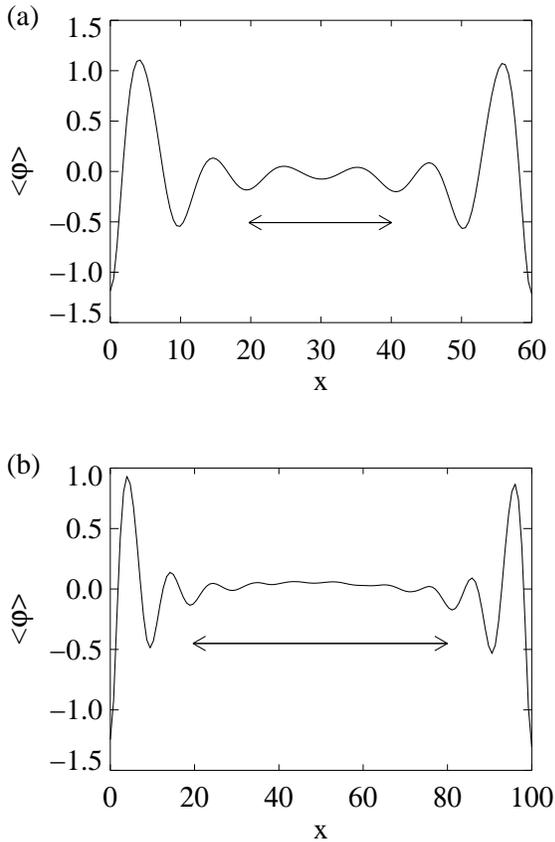} 
\end{center}
\vspace{-.5cm}
\caption{\label{twosizes} Average patterns of the front $\varphi$ for the 
one-dimensional
Kuramoto-Sivashinsky equation with stress-free boundary conditions.
a) $L=60$. b) $L=100$. The arrows indicate the `central region', as defined in
the text. }
\end{figure}

\vspace{-1cm}
\begin{figure} 
\begin{center}
\def\epsfsize#1#2{0.39\textwidth}
\leavevmode
\epsffile{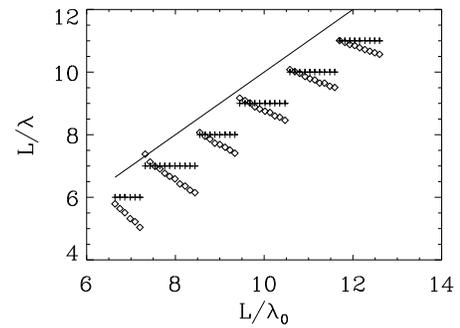} 
\caption{\label{wavenumbers} +: The number of maxima $N=L/\lambda$ in the 
average front versus the aspect ratio $L/\lambda_0$, $\lambda$ 
being the average
distance between consecutive minima in the entire region. $\diamond$: 
the number of maxima in the central region, given by $L/\lambda_c$. 
The solid line corresponds to $\lambda=\lambda_0$.}
\end{center}
\end{figure}

\vspace{-1cm}

\begin{figure} 
\begin{center}
\def\epsfsize#1#2{0.38\textwidth}
\leavevmode
\epsffile{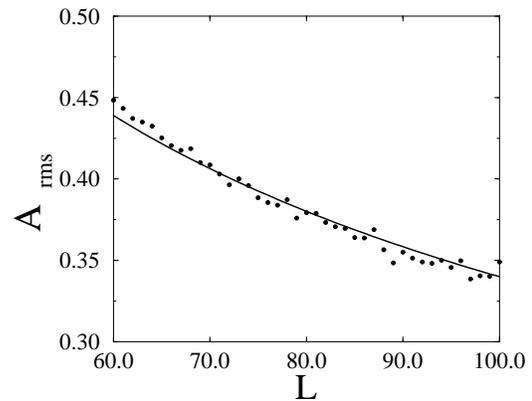} 
\end{center}
\caption{\label{Arms} Variation of the root mean square amplitude $A_{rms}$ 
over the range of system sizes $59<L<107$. The solid curve is the
function $C/\sqrt{L}$, with $C$ fitted to the data}
\end{figure}

\end{twocolumns}
\end{document}